\documentstyle[amstex,amssymb,11pt]{article}

\begin{document}

\title{{\bf Power law distribution in High School Education: Effect \ of
Economical, Teaching and study conditions.}}
\author{Hari M. Gupta, Jos\'{e} R. Campanha and F\'{a}bio R. Chavarette \\
Departamento de F\'{\i}sica- IGCE\\
and Virtual Research Center - Science and Computing for Complexity
(CC@Complex\\
Universidade Estadual Paulista (UNESP)\\
C.P. 178, Rio Claro 13500-970, SP, Brazil\\
\\
email - campanha@rc.unesp.br}
\date{}

\begin{abstract}
We studied the statistical distribution of student\'{}s performance, which
is measured through their marks, in university entrance examination
(Vestibular) of UNESP (Universidade Estadual Paulista) with respect to (i) \
period \ of study - day vs. night period (ii) teaching conditions - private
vs. public school (iii) economical conditions - high vs. low family income.
This examination reflect quality of high schools education. We observed long
ubiquitous power law tails in Physical and Biological sciences in all cases.
The mean value increases with better study conditions followed by better
teaching and economical conditions. In humanities, the distribution is close
to normal distribution with very small tail. This indicate that these power
law tail in science subjects are due to the nature of the subject itself.
Further better study, teaching and economical conditions are more important
for physical and biological sciences in comparison to humanities. We explain
these statistical distributions through Gradually Truncated Power law
distributions. We discuss the possible reason for this peculiar behaviour
and make suggestions to improve science education at high school level.


\end{abstract}

\newpage

{\bf I. Introduction}

\medskip

Power law distribution [1-5] has been first noted by Pareto in economics [6]
in 1897, and afterward, by others in many physical [7-10], biological
[11,12], economical [13-17] and more recently in educational [18] complex
systems. Recently, we studied the statistical distribution of student\'{}s
performance, which is measured through their marks, in university entrance
examination (Vestibular) of UNESP (Universidade Estadual Paulista) for years
1998, 1999, and 2000. We observed long ubiquitous power law tails in
physical and biological sciences as have been observed in many physical and
economical systems. We explain these statistical distributions through
Gradually Truncated Power law distributions which we defined in line with
Gradually Truncated L\`{e}vy Flight [19, 20] for stochastic processes. In
humanities we have almost normal distribution.

We postulate that these long ubiquitous power law tails may be due to the
nature of subject itself and/or due to better economical, teaching and
studying conditions. In the present paper we study the statistical behaviour
of candidates marks in physical, biological and humanities with respect to
candidate\'{}s economical, teaching and studying conditions for the
University Entrance Examination (Vestibular) of UNESP (Universidade Estadual
Paulista) - S\~{a}o Paulo- Brazil. We found that power law tails are present
in physical and biological sciences in all cases. The mean value increases
more rapidly with better studying conditions followed by better teaching
conditions and economical conditions. In Humanities the statistical
behaviour is approximately normal distribution with \ very small tails. The
mean value increases in the same way as observed in physical and biological
sciences but in small magnitude.

This study is interesting as treat educational system as complex system and
bring out relative importance of different factors on education and peculiar
nature of mathematics and science subjects. This in turn can help to improve
science education at high school.

\medskip

\smallskip {\bf II. Data Analysis}

\medskip

Pareto [6] proposed a power law distribution based on positive feedback. He
found that the distribution of income is well approximated by an inverse
power law. This feedback effect decreases gradually after certain step size
due to physical limitation of the components of the system. The whole
process can be described through Gradually Truncated power law distribution
given as follow [19,20]:

\smallskip

\begin{equation}
P(x)=\frac{c_{1}}{c_{2}+(\left| x-x_{m}\right| )^{1+\alpha }}f(x)
\end{equation}

\smallskip

\noindent where $P(x)$ is the probability of taking a step of size $x$, $%
x_{m}$ is the value of $x$ for maximum probability, $c_{1}$ and $c_{2}$ are
constants and are related through:

\smallskip

\begin{equation}
c_{1}=c_{2}P(x_{m})
\end{equation}

\smallskip

\noindent $c_{2}$ can be obtained through normalization condition. Further

\smallskip

\begin{equation}
f(x)=\left\{
\begin{array}{ccl}
1 & \mbox{if} & \left| x\right| \leqslant x_{c} \\
\exp \left\{ -\left( \frac{\left| x\right| -x_{c})}{k}\right) ^{\beta
}\right\} & \mbox{if} & \left| x\right| >x_{c}
\end{array}
\right.
\end{equation}

\smallskip

\noindent where $x_{c}$ $(x_{c}>x_{m})$ is the critical value of step size,
where the probability distribution began to deviate from power law
distribution due to physical limitation, $k$ gives the sharpness of the
cut-off. It is a general tendency in these systems to approach a normal
distribution in the large scale and thus we choose:

\smallskip

\begin{equation}
\beta =2-\alpha
\end{equation}

\smallskip

\noindent $(1+\alpha )$ is the power of the power law distribution.

The normal distribution [21] is given by:

\smallskip

\begin{equation}
P(x)=\frac{1}{\sqrt{2\pi }\sigma }\exp [-\frac{(x-\stackrel{\_}{x})^{2}}{%
2\sigma ^{2}}]
\end{equation}

\noindent where $\stackrel{\_}{x}$ and $\sigma $ are respectively mean value
and standard deviation.

In the present paper, we analyze data of marks obtained by candidates in
University Entrance examination of Universidade Estadual Paulista (UNESP) in
state of S\~{a}o Paulo - Brazil, in years 1998, 1999 and 2000 [22]. About
sixty thousand candidates appear each year in this examination. This
examination is divided in three groups depending on the course chosen by the
candidate. These groups are i) Physical sciences, ii) Biological sciences
and iii) Humanities. Each candidate takes three examinations: General
Knowledge, Specific Knowledge of the area and Portuguese Language. General
Knowledge and Portuguese Language examinations are the same for all
candidates, while the examination of Specific Knowledge is different and is
on the area chosen.

In our earlier paper [18], we observed peculiar behaviour in marks obtained
by candidates in the examination of specific knowledge. We therefore, in the
present paper are considering statistical distribution of candidates marks
only in this examination. To see the effect of various factors on the
evolution of student\'{}s knowledge, we compare the statistical distribution
in physical sciences, biological sciences and humanities for: i) Period of
study - Day time vs. Night time students, ii) Private vs. Public schools and
iii) High vs. low family income.

In Figure 1 , we compare the marks obtained by students studying in day time
and night times for the year 2000. Generally night time students work during
the day and thus have less time to study. We compare physical sciences,
biological sciences and humanities students in Figures 1a, 1b and 1c
respectively. In physical and biological sciences, the distribution is given
by gradually truncated power law, while in humanities it approaches to
normal distribution with small tail.

In Figure 2 we compare the marks obtained by students studying in private
and public schol in physical sciences (Figure 2a), biological sciences
(Figure 2b) and humanities (Figure 2c). Normally teaching conditions are
better in private schools because in Brazil, the existence of these schools
depend on the performance of their students in these types of examinations.
Here again the distribution is given by gradually truncated power law in
physical and biological sciences while normal distribution with small tail
in humanities.

In Figure 3 we compare the marks obtained by students coming from high
income group, i.e. total family income more than ten minimum salary ($%
\approx $US 800.00 dollars/month), and low income group, i.e. total family
income less than ten minimum salary. In Brazil, families with income above
than ten minimum salary normally have appropriate conditions to live and
study. Again in this case distributions are the same as observed earlier.

The parameters of the distribution for drawing theoretical curves for
physical and biological scieces are given in Table 1 and 2 respectively. The
value of $(1+\alpha )$ is between 1 and 2 in all cases, thereby giving long
distribution tail. The values of x$_{C}$ is around 80 in all cases, where
positive feedback began to decrease gradually, i.e. probability decreases
more rapidly than given by power law. The agreement of theoretical curves is
good with empirical curves in all cases.

\bigskip

\begin{center}
Table 1 - Physical Sciences

\smallskip
\begin{tabular}{||c||c||c||c||c||c||}
\hline\hline
& $(1+\alpha )$ & $x_{m}$ & $x_{c}$ & $k$ & $\stackrel{\_}{x}$ \\
\hline\hline
Day time & 1.11 & 6 & 83 & 200 & 31.6 \\ \hline\hline
Night time & 1.47 & 4 & - & - & 13.2 \\ \hline\hline
Private School & 1.19 & 7 & 75 & 200 & 37.9 \\ \hline\hline
Public School & 1.41 & 4 & - & - & 17.5 \\ \hline\hline
High income & 1.0 & 6 & 77 & 200 & 34.2 \\ \hline\hline
Low income & 1.33 & 4 & 87 & 227 & 18.9 \\ \hline\hline
\end{tabular}

\bigskip Table 2 - Biological Sciences

\smallskip
\begin{tabular}{||l||l||l||l||l||l||}
\hline\hline
& $(1+\alpha )$ & $x_{m}$ & $x_{c}$ & $k$ & $\stackrel{\_}{x}$ \\
\hline\hline
Day time & 1.35 & 6 & 82 & 62 & 32.3 \\ \hline\hline
Night time & 1.80 & 2 & - & - & 12.6 \\ \hline\hline
Private School & 1.51 & 14 & 85 & 23 & 37.6 \\ \hline\hline
Public School & 1.53 & 4 & 91 & 14 & 16.6 \\ \hline\hline
High income & 1.42 & 5 & 82 & 50 & 34.4 \\ \hline\hline
Low income & 1.63 & 2 & 83 & 27 & 20.6 \\ \hline\hline
\end{tabular}

\smallskip Table 3 - Humanities

{\bf \bigskip }
\begin{tabular}{||l||l||l||}
\hline\hline
& $\overline{x}$ & $\sigma $ \\ \hline\hline
Day time & 43.4 & 15.5 \\ \hline\hline
Night time & 31.8 & 14.5 \\ \hline\hline
Private School & 47.7 & 14.7 \\ \hline\hline
Public School & 35.2 & 14.9 \\ \hline\hline
High income & 45.6 & 15.4 \\ \hline\hline
Low income & 35.1 & 15.1 \\ \hline\hline
\end{tabular}
\end{center}

We further observe that all the curves of different years collapse in a
single curve so far as general behaviour is concerned [23]. A small
variation in the magnitude is due to the relative variation of the
dificulties felt by the candidates in answering the questions in the
examination, from one year to another year. The basic facts of the process
of teaching and evaluation remain the same.

\medskip

\smallskip {\bf III. Discussion}

\medskip

In general, the educational system is very complex. As this examination
involves students who studied in schools with different facilities,
traditions and teaching levels and come from families of different social,
racial and economical groups, which in turn provide different financial and
emotional conditions, the problem becomes still more complex. Broadly
speaking, we can say tthe following:

(i) The better study conditions, class room teaching and economical
facilities improve the performance of students in all the areas. However in
physical and biological science group (physics, chemistry, mathematics and
biology) it is more important than humanities. We found that the average
marks obtained by day time students are better than night time students by
140\% in physical sciences, 157\% in biological sciences and 37\% in
humanities. Further the average marks obtained by private school students
are better than public school students by 117\% in physical sciences, 126\%
in biological sciences and 36\% in humanities. Finally, the average marks
obtained by high income group students compare to low income group students
are better by 81\% in physical sciences, 67\% in biological sciences and
27\% in humanities.

(ii) Better study conditions, are more important followed by teaching and
economical conditions.

(iii) We observed long power law distribution tail in physical and
biological sciences in all group. This shows that this peculiar behaviour is
due to nature of the subject itself. In science and mathematics, all topics
are inter-related. To understand a topic, a student need to know the earlier
topics given. This creat a long term memory effect thereby giving power law
effect. In case of humanities, topics are independent and in most cases one
can understand a topic without knowing a topic given earlier particularly at
this level of education. This gives normal distribution.

As far as we know, this kind of study is just begining in the field of
education. It is therefore not possible for us to speculate some definite
reasons for this kind of behaviour. However, we feel that similar kind of
studies with different examinations, like school leaving certificate
examination etc., in different countries with different social, economical
and educational structures and social inhomogenity, can provide a better
clarification of the relative importance of each factor in the preparation
of a student in a particular area of study.

In view of our observations, we make the following suggestions to improve
science and mathematics education at high school level:

{\bf A)} Scholarship should be given to talented poor student in science and
mathematics at high school level. This will give them sufficient time to
study. The student can be selected through a appropiate state or national
examination.

{\bf B)} Generally students feel science and mathematics as some thing
abstract not related to our day life. Good science laboratories are a
necessary part of science education [24] and make science ducation more
interesting.

{\bf C)} In schools, facilities should be provided for students to resolve
their problems in science subjects. It can help them to have good idea of
basic concepts. Perhaps tutorial is a good idea for this purpose.

\begin{center}
{\bf Figure Captions}

\bigskip
\end{center}

\medskip

\noindent Figure {\bf 1: }Number of candidates obtaining marks $x$ $(N(x))$ $%
vs.$ marks obtained $(x)$ for specific knowledge examination for year 2000
among day time and night students. Figure 1a is for Physical science
subjects; Figure 1b is for Biological science subjects and Figure 1c is for
Humanities subjects. The theoretical curves are through Gradually Truncated
power law distribution in (a) and (b), while through normal distribution in
(c).

\medskip

\noindent Figure {\bf 2}: Number of candidates obtaining marks $x$ $(N(x))$ $%
vs.$ marks obtained $(x)$ for specific knowledge examination for year 2000
among private and public schools. Figure 2a is for Physical science
subjects; Figure 2b is for Biological science subjects and Figure 2c is for
Humanities subjects. The theoretical curves are through Gradually Truncated
power law distribution in (a) and (b), while through normal distribution in
(c).

\bigskip

\noindent Figure {\bf 3}: Number of candidates obtaining marks $x$ $(N(x))$ $%
vs.$ marks obtained $(x)$ for specific knowledge examination for year 2000
among high and low income family students. Figure 3a is for Physical science
subjects; Figure 3b is for Biological science subjects and Figure 3c is for
Humanities subjects. The theoretical curves are through Gradually Truncated
power law distribution in (a) and (b), while through normal distribution in
(c).


\begin{thebibliography}{99}
\bibitem{}  Vestibular Report - We are thankfull to VUNESP for providing
vestibular results to analyse the data
\end{thebibliography}
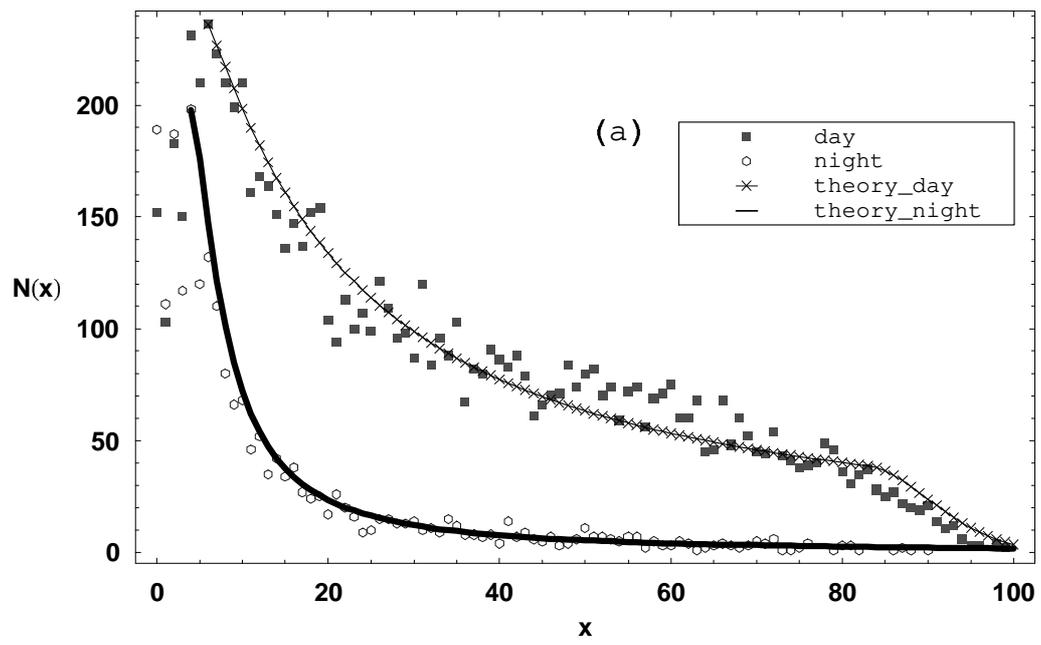
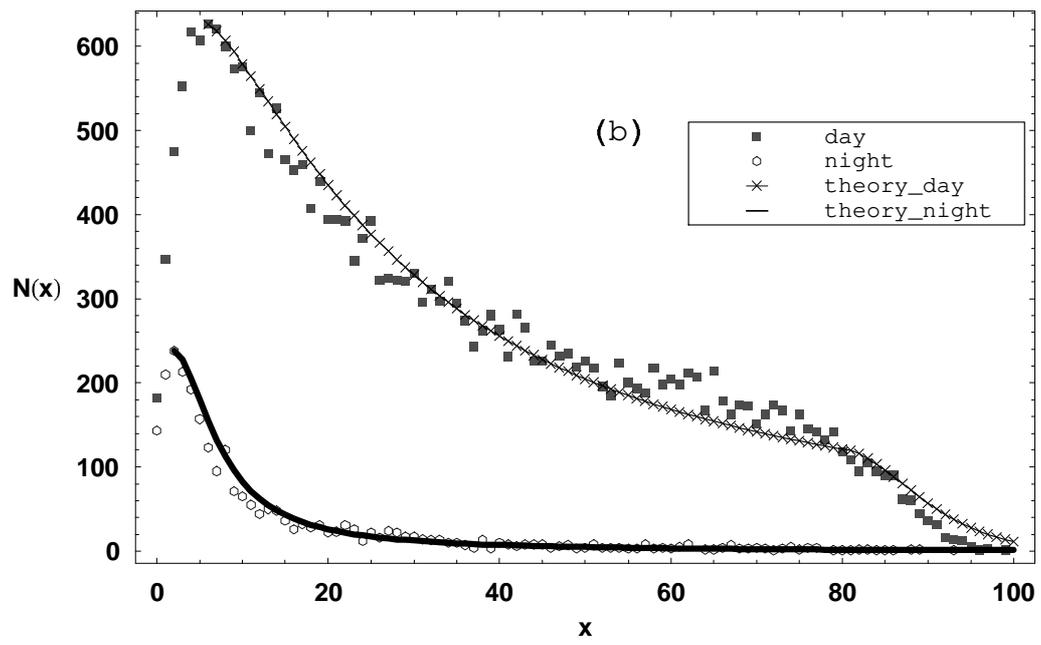
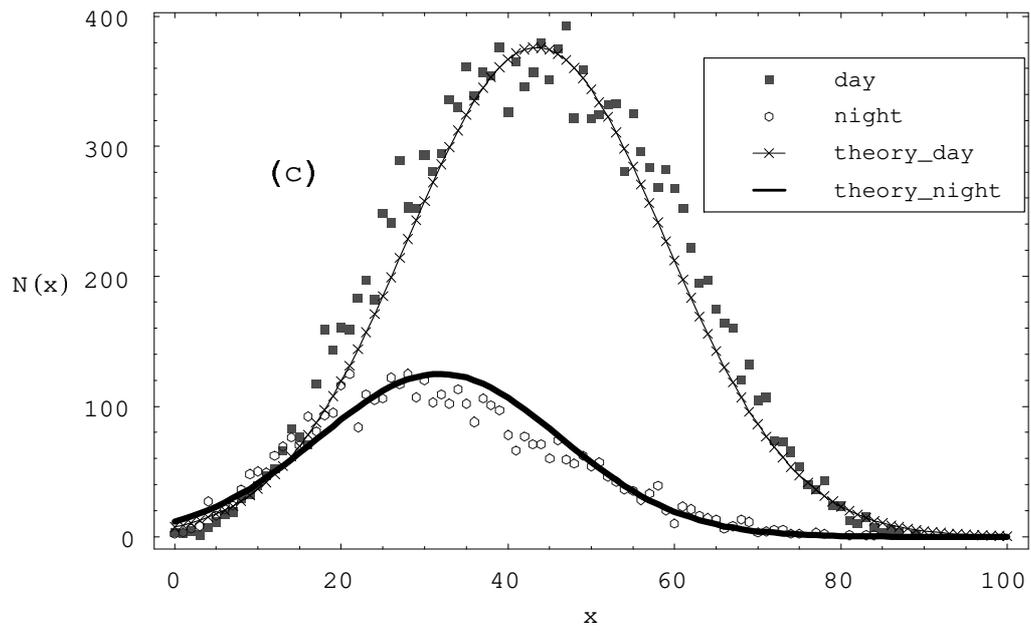
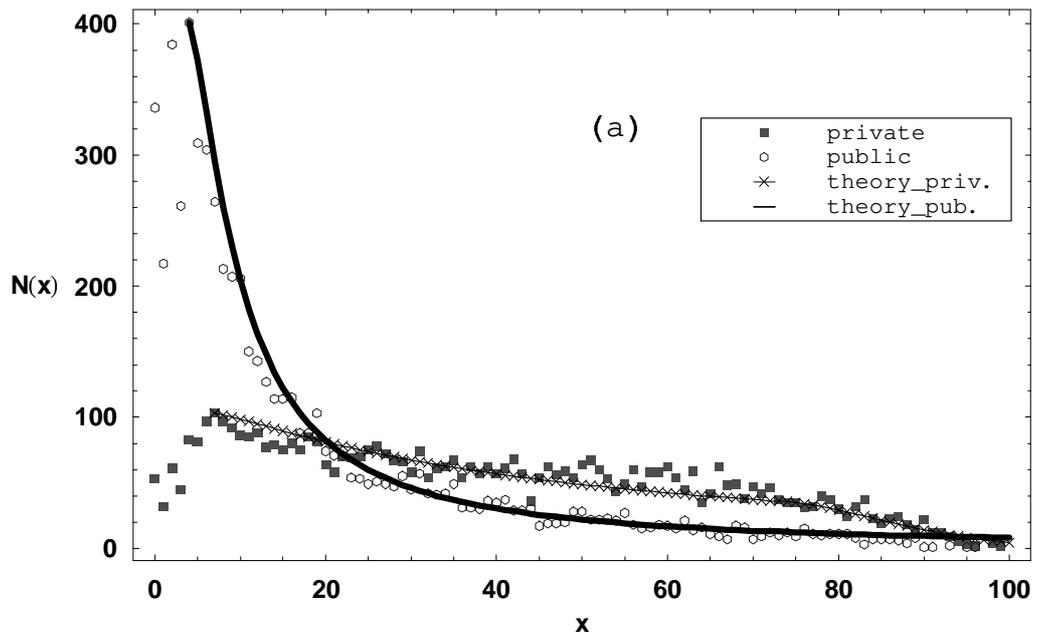
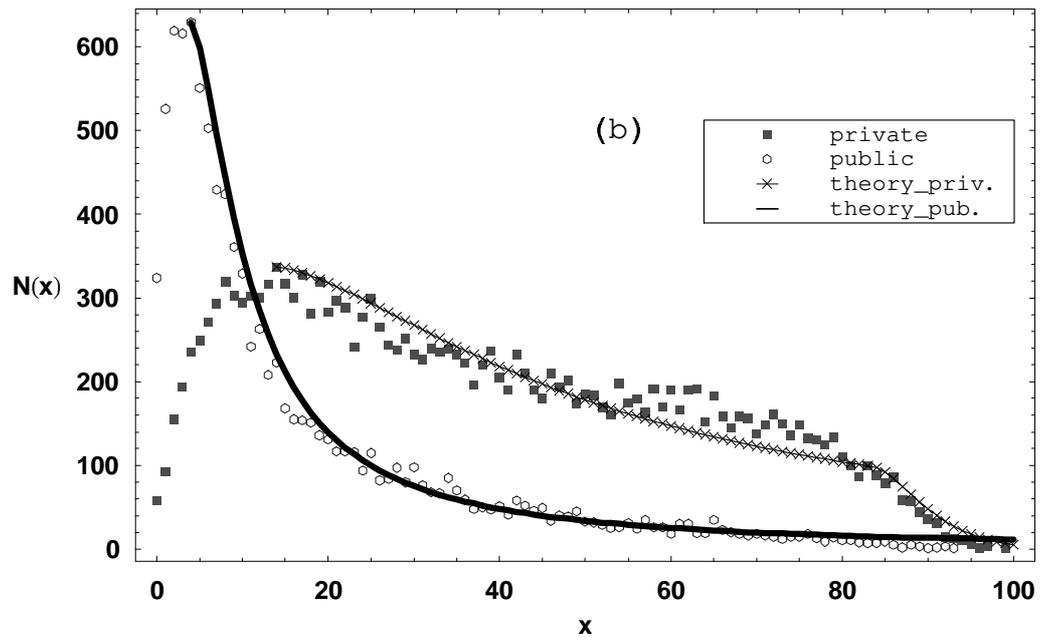
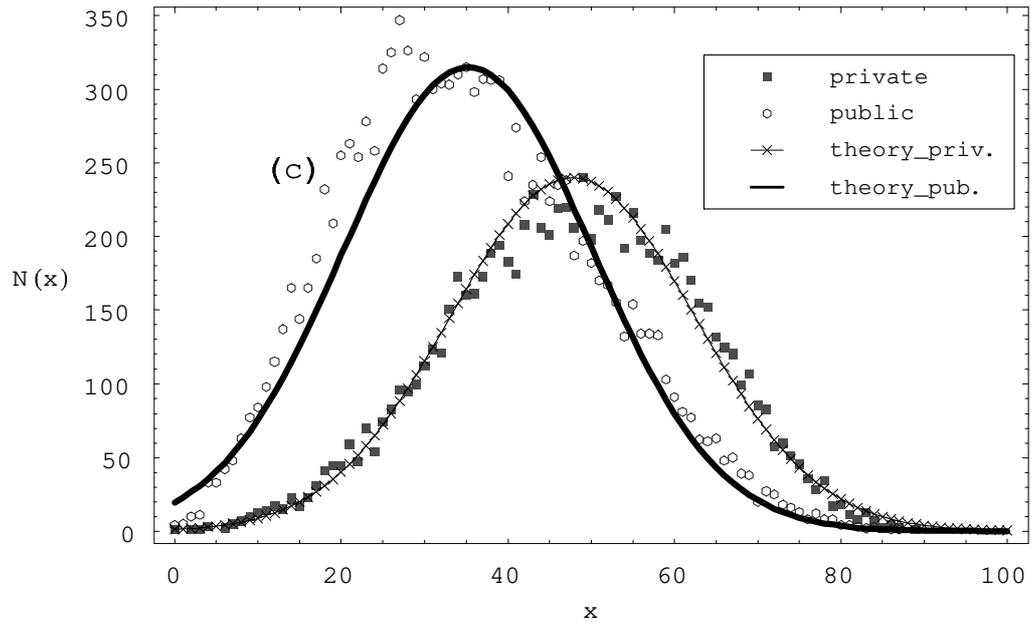
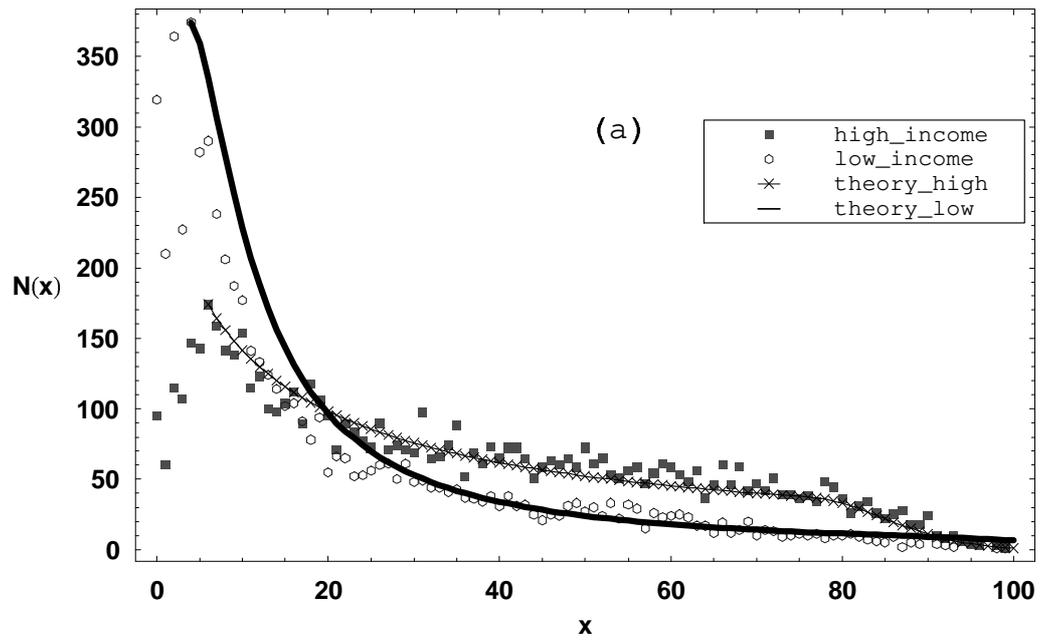
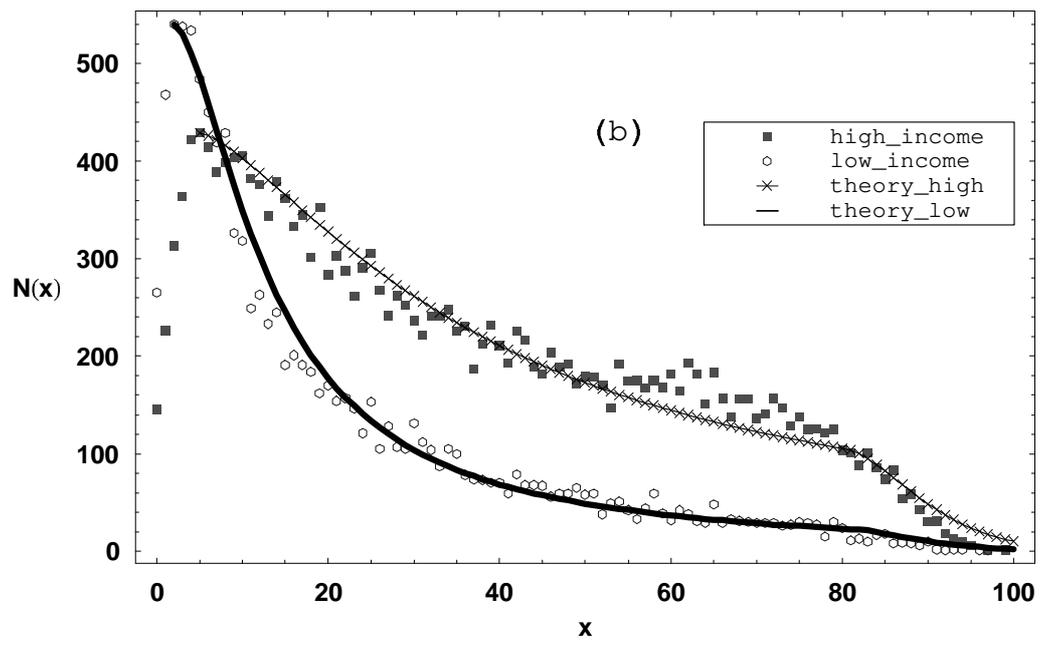
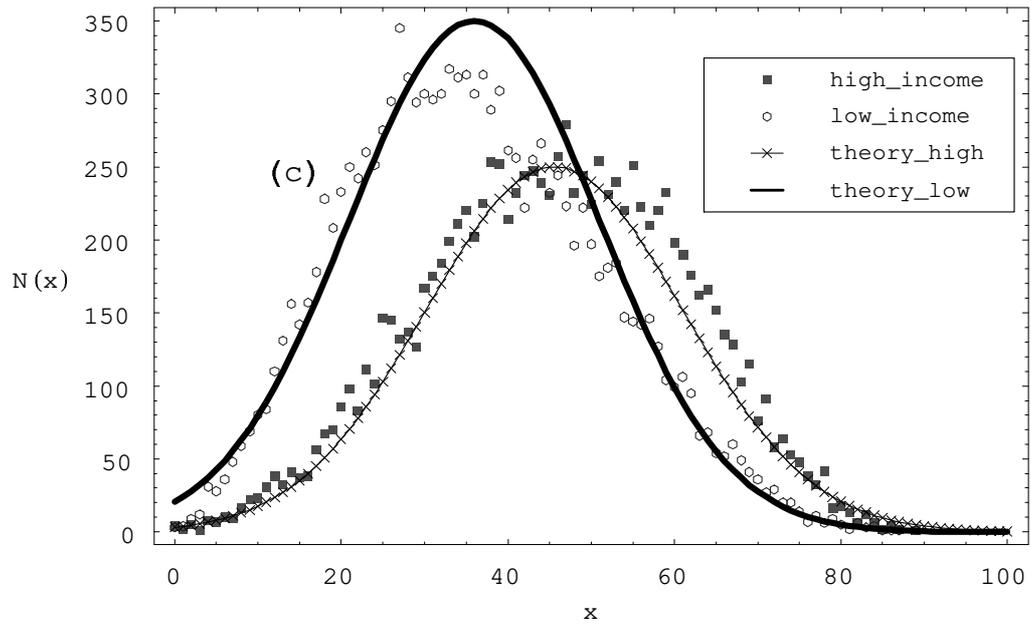
\end{document}